\documentclass[graybox]{svmult}

\usepackage{amsmath,amssymb}
\usepackage[all]{xy}
\usepackage{graphicx}

\usepackage{hyperref}

\DeclareMathOperator{\Li}{Li}
\DeclareMathOperator{\tr}{tr}
\DeclareMathOperator{\sgn}{sgn}
\DeclareMathOperator{\Alt}{Alt}

\newcommand{\grp}[1]{\mathrm{#1}}

\graphicspath{{fig/}}

\begin{document}

\title{Polylogarithm identities, cluster algebras and the \texorpdfstring{\(\mathcal{N} = 4\)}{N=4} supersymmetric theory\thanks{Expanded version of a talk given at the Opening Workshop of the Research Trimester on Multiple Zeta Values, Multiple Polylogarithms, and Quantum Field Theory, organized by Jos\'e I.~Burgos Gil, Kurusch Ebrahimi-Fard, D.~Ellwood, Ulf K\"uhn, Dominique Manchon and P.~Tempesta.}}

\author{Cristian Vergu}

\institute{%
Cristian Vergu \at%
Department of Mathematics, King's College London\\
The Strand, WC2R 2LS, London, UK\\
\email{c.vergu@gmail.com}%
}
\maketitle

\abstract{%
Scattering amplitudes in \(\mathcal{N} = 4\) super-Yang Mills theory can be computed to higher perturbative orders than in any other four-dimensional quantum field theory.  The results are interesting transcendental functions.  By a hidden symmetry (dual conformal symmetry) the arguments of these functions have a geometric interpretation in terms of configurations of points in \(\mathbb{CP}^3\) and they turn out to be cluster coordinates.  We briefly introduce cluster algebras and discuss their Poisson structure and the Sklyanin bracket.  Finally, we present a \(40\)-term trilogarithm identity which was discovered by accident while studying the physical results.
}

\section{Introduction}
\label{sec:introduction}

There is no doubt that quantum field theory and mathematics are deeply connected.  There are many examples where field theory intuition helped formulate mathematical conjectures or even theorems (Seiberg-Witten theory in topology~\cite{Witten1994}, Wilson loops in Chern-Simons theory for knot theory~\cite{Witten1989}).  Similarly, progress in mathematics has stimulated progress in field theory (as a prime example we have ADHM construction~\cite{Atiyah1978} of instantons, but also work in index theory~\cite{Atiyah1963} which helped in the understanding of field theory anomalies).  And these are just a few of many examples.

In this review we will focus on one of the many connecting bridges between quantum field theory and number theory: polylogarithms.  In quantum field theory polylogarithms and the closely related multiple zeta values are ubiquitous.  They arise in the perturbative computations of various quantities.

There are many quantities one may attempt to compute and, moreover, there are many different quantum field theories.  Many results are already available but frequently the complexity of the final answers (not to mention the complexity of the computation) is forbidding.  We are then naturally led to ask which field theories and what quantities are most likely to be understood in simple terms.

These questions, while very natural, are not at all obvious, but in recent years an answer has began to emerge.  As we will explain, the answer is somewhat surprising.  The textbook example for the simplest interacting field theory is called the \(\phi^4\) theory.  This is a theory of a single scalar field with a four-point interaction.  The Feynman diagrams in this theory have internal vertices of degree four. Many results are known in this theory see, for example, ref.~\cite{BROADHURST1995, Schnetz2010}.  However, it has recently emerged that there is a better candidate for study, which we will discuss below.

Relativistic field theories are symmetric under the Poincar\'e group. The Poincar\'e group has the Lorentz group \(\grp{O}(1,3)\) as a subgroup and particles are in correspondence with irreducible representations of these symmetry groups.  The scalar particles transform in the trivial representation of \(\grp{O}(1,3)\) so they realize the relativistic symmetry in the simplest possible way.  As mentioned above, the \(\phi^4\) theory is a theory of scalar (or spin zero) fields.

Other representations of the Lorentz symmetry may appear: fermions which transform as a representation of the covering group \(\grp{Spin}(1,3)\), gauge fields which are vectors of \(\grp{O}(1,3)\), the graviton which is rank two tensor representation, etc.  In the case of the gauge fields and of the graviton the formulation of the quantum theory is complicated by the fact that states are defined modulo gauge transformations.  This also complicates the computations since one has to make a choice of gauge (or a choice of representative in the equivalence class).

Despite these technical complications, in many cases the final results, when expressed in terms of appropriate variables, turn out to be strikingly simple (the computation of Parke and Taylor in ref.~\cite{PhysRevLett.56.2459} being a prime example).  Then, we are led to suspect that there should be more efficient ways to find these answers.

We have briefly discussed the theories but we still haven't specified the types of quantities we are going to compute.  We turn to this question next.  The quantities which will be most relevant in the following discussion are scattering amplitudes.  Let us give a rough definition of scattering amplitudes.  A field theory of the kind we will consider is defined by a functional \(S[\phi]\) called action, depending of functions \(\phi(\vec{x},t)\) called fields (here \(t\) is time, \(\vec{x}\) is a three-dimensional vector and \(\phi\) is a generic name for a field; in general the theory can contain several fields with different \(\grp{O}(1,3)\) transformations).  From this functional we can obtain by variational methods partial differential equations (called equations of motion) for the fields of the theory. Now, given some boundary conditions \(\phi_{\pm}\) at \(t = \pm \infty\) for the fields, from the solution \(\phi_0\) to the equation of motion satisfying these boundary conditions one can build a complex number \(\exp(i S[\phi_0])\) which is called the tree level amplitude of transition between \(\phi_{-}\) and \(\phi_{+}\) (if there is no solution for the prescribed boundary conditions, then the amplitude is defined to be zero).  The name `tree' is due to the fact that this quantity can be computed as a sum of tree-shaped Feynman diagrams.

The computation using the definition can be tedious in general, especially for gauge theories where one has to make an arbitrary choice of gauge (in the final result the dependence on this arbitrary choice must cancel; when this happens we call the answer `gauge invariant').  The tree level amplitudes have two important properties: analyticity (in a certain domain) and factorization.\footnote{Analyticity survives after adding quantum corrections, but factorization becomes more subtle in case there are infrared divergences (see ref.~\cite{Bern1995}).  Since scattering amplitudes in gauge theories are infrared divergent, exploiting factorization at loop level seems to be much harder.}  Factorization here means that the amplitude has certain poles whose residues are products of simpler amplitudes.  The requirement of factorization is a very powerful constraint; using it, the BCFW (\cite{Britto2005}) recursion relations allow the computation of all tree-level amplitudes of the \(\mathcal{N} = 4\) theory we will describe in the next section.

In the quantum theory graphs with loops appear as well.  Graphs with loops correspond to non-trivial integrals, which yield mathematically interesting results.  It is an empirical observation that the transcendentality of an \(\ell\)-loop result is bounded from above by \(2 \ell\); for a one-loop quantity the most complicated part can be expressed in terms of dilogarithms.

For theories relevant experimentally, like Quantum Chromodynamics (QCD), a one-loop answer will contain not only dilogarithms, but also logarithms and even rational terms.  The transcendentality of the answer is not uniform.  However, for the special case of \(\mathcal{N} = 4\), the answers are of uniform transcendentality.  In some cases, see ref.~\cite{Kotikov2004}, the \(\mathcal{N} = 4\) answer can be obtained from the uniform transcendentality of the more complicated QCD result.

\section{The maximally supersymmetric theory}

We mentioned previously that the theories with spin are in some sense simpler than theories of scalar (spinless) particles.  Even so, there are many possible theories of particles with spin.  Supersymmetry is a remarkable symmetry which can transform between particles of different spins.  The maximal supersymmetry of a non-gravitational theory in three space and one time dimensions is called \(\mathcal{N} = 4\) supersymmetry.  The reason for the name is that \(\mathcal{N} = 1\) supersymmetry is the minimal supersymmetry and the maximal supersymmetry has four times as many supersymmetries as the minimal one.

In ref.~\cite{Coleman1967}, Coleman and Mandula proved a theorem about the possible symmetries of a relativistic theory.  Under certain assumptions they showed that the symmetry group has the structure of a product between the Lorentz and some other `internal' symmetry group. Later, Haag, \L opusza\'nski and Sohnius~\cite{Haag1975} showed that a non-trivial symmetry structure is possible, but it has to be a supergroup symmetry, not a Lie group symmetry.  A supergroup is obtained by exponentiating Lie superalgebra elements, where a Lie superalgebra is a \(\mathbb{Z}_2\)-graded algebra with a bracket satisfying graded commutativity and a graded version of Jacobi identity.  The supergroup has a usual Lie group as a subgroup and, somewhat surprisingly, this is also enlarged with respect to a typical relativistic theory.  In a relativistic theory the symmetry group is the Poincar\'e group, which now gets enhanced to a \(\grp{SO}(2,4)\) group, also known as the conformal group.  The new symmetries are the dilatation \(D\) and four conformal transformations \(K_{0}, \dots, K_{3}\).

The theory with maximal supersymmetry was constructed shortly after in ref.~\cite{Brink1977} by Brink, Schwarz and Scherk.  This theory is uniquely defined by its symmetry.  It is a theory of a connection \(A\) on an \(\grp{SU}(N)\) principal bundle over Minkowski space \(\mathbb{M}\), together with fermionic field \(\Psi\) and scalar fields \(\Phi\).  The action functional is given by the Yang-Mills term together with other terms dictated by supersymmetry, which we do not write explicitly since they will not be important in the following
\begin{equation}
  S[A, \Psi, \Phi] = \frac 1 {2 g^2} \int_{\mathbb{M}} \tr(F \wedge * F + \dots).
\end{equation}
Here the trace is taken in the fundamental representation of \(\grp{SU}(N)\) and \(g^2\) is a real number, called coupling constant.  \(F = d A + A \wedge A\) is the curvature of the connection \(A\) and \(* F\) is the Hodge dual.  The scattering amplitudes, can be expanded as a power series in \(g\).

Terms in the perturbative expansion are computed by summing Feynman graphs.  The contribution of a Feynman graph can be factored in two different types of terms: the kinematic part, depending on the positions (or on the momenta after Fourier transform) and the `color' part which depends on the Lie algebra \(\mathfrak{su}(N)\) of the gauge group \(\grp{SU}(N)\).  The observables can then be decomposed on a basis of \(\mathfrak{su}(N)\) invariants whose coefficients depend on \(N\) and \(g\).  If we select invariants which can be written as a single trace and, for these terms, we select the dominant behavior when \(N \to \infty\), then the topology of the contributing graphs simplifies.  We find that only planar graphs contribute.  The way to select the planar graph contributions is to reorganize the perturbation theory as an expansion in \(\lambda = g^2 N\) around \(\lambda = 0\), with \(N \to \infty\) and \(g^2 \to 0\).  This is the well-known 't Hooft limit~\cite{Hooft1974}.

From his study of the large \(N\) limit, 't Hooft conjectured that the result in the 't Hooft limit is the genus zero term in an expansion of a theory which sums over surfaces.  A theory which sums over surfaces is a string theory (in a theory of particles, one sums\footnote{The sum over particle histories is not well-defined mathematically. Nevertheless, we can use it formally to compute the perturbative expansion.  A similar statement holds for a string theory, where we sum over string histories also called worldsheets.} over particle paths, as instructed by the Feynman path integral).  The conjecture also stated that subleading terms in \(N\) correspond to sums over surfaces of higher genera.

This conjecture of 't Hooft is very general, and was initially proposed for QCD, where the gauge group \(\grp{SU}(3)\) was to be replaced by \(\grp{SU}(N)\).  It was hoped that understanding \(N \to \infty\) case could shed some light on the \(N=3\) case.  If instead of QCD we consider the \(\mathcal{N}=4\) supersymmetric theory, the conjecture was sharpened by the AdS/CFT correspondence of Maldacena (see ref.~\cite{Maldacena1999}).  The AdS/CFT correspondence identifies the precise measure on the space of surfaces.  In fact, we should use super-strings, but if we set the fermions to zero we obtain a theory of a string moving in an \(\text{AdS}_5 \times \mathbb{S}^5\) geometry.  Here CFT means Conformal Field Theory, which in this case is a theory with a symmetry group containing \(\grp{SO}(2,4)\).  The \(\text{AdS}_5\) space is the five-dimensional hyperbolic space with a non-definite metric, which can be obtained by analytically continuing some coordinates to imaginary values (a procedure called Wick rotation in the Quantum Field Theory literature).  This is similar to the relation between Euclidean space \(\mathbb{R}^4\) and Minkowski space \(\mathbb{M}\).  The isometry group of \(\text{AdS}_5\) is again \(\grp{SO}(2,4)\).  In fact, the full \(\grp{PSU}(2,2\vert 4)\) symmetry groups match on both sides of the correspondence.

The AdS/CFT duality describes a physical system in two different ways. When the 't Hooft coupling \(\lambda\) is small, the field theory perturbative expansion in powers of \(\lambda\) is reliable.  When the 't Hooft coupling is large, instead, one should use string theory on the \(\text{AdS}_5 \times \mathbb{S}^5\) should be used instead.  In this case, the expansion variable is \(\lambda^{-1/2}\).  Therefore, the duality is of strong-weak type; the strong coupling (\(\lambda \to \infty\)) in the CFT can be mapped to a weakly coupled description in the dual string theory.

The computation of the scattering amplitudes can also be done in the dual string theory, as described in ref.~\cite{Alday2007a}.  In the dual string theory scattering amplitudes are given by the exponential of a minimal surface in \(\text{AdS}_5\) which ends on the boundary of \(\text{AdS}_5\) on a polygon whose sides are the momenta of the scattered particles (the polygon closes by momentum conservation).

\section{Kinematics}
\label{sec:kinematics}

In this section we describe the kinematics of a scattering process in terms of configurations of points in \(\mathbb{CP}^3\).  This was initiated in ref.~\cite{Hodges:2009hk} for tree-level amplitudes, later extended to superspace in ref.~\cite{Mason2009a} and further studied in ref.~\cite{ArkaniHamed:2009dn}.  The usefulness of these variables for loop amplitudes was emphasized in ref.~\cite{Arkani-Hamed2011} and also in ref.~\cite{Goncharov2010} for an explicit two-loop result.

Consider an \(n\)-particle scattering process.  The particle labeled by \(i\) is described by the on-shell momentum \(p_{i}\) (with \(p_{i}^{2} = 0\), where the norm is computed using the Minkowski metric), its helicity \(s_i\) and a gauge algebra generator \(t_{i} \in \mathfrak{su}(N)\).  The helicity labels the representation under the compact subgroup \(\grp{U}(1)\) of the Lorentz group \(\grp{O}(1,3)\) which preserves the momentum \(p_i\).  In fact, if our theory contains fermions we need to pass to the covering group \(\grp{Spin}(1,3)\) of part of the Lorentz group connected to the identity.   In the end, the representations turn out to be labeled by \(s \in \mathbb{Z}/2\).

As we discussed above, in the 't Hooft limit \(N \to \infty\), \(g^{2} N = \lambda\) fixed, only single-trace terms survive in the scattering amplitudes.  If we look at one of these single-trace terms, we see that the scattered particles are cyclically ordered.  We can therefore introduce a \emph{dual} space with coordinates \(x\) such that the momenta \(p_{i}\) are expressed as \(p_{i} = x_{i-1} - x_{i}\).  The \(x_i\) coordinates are only defined up to a translation \(x_i \sim x_i + a\).  We denote by \(\tilde{\mathbb{M}}\) the space parametrized by dual coordinates \(x\).

The \(\mathcal{N}=4\) super-Yang-Mills theory is superconformal invariant.  Besides this superconformal symmetry, the \(\mathcal{N}=4\) super-Yang-Mills theory also has a surprising \emph{dual} superconformal symmetry, whose bosonic subgroup acts on the dual coordinates \(x\).  In the following we will mostly be interested in the conformal subgroup of this dual superconformal group.  The dual superconformal symmetry is a hidden symmetry, which only arises in the 't Hooft limit.  In particular, it can not be verified on the Lagrangian of the theory.

Historically, this symmetry arose as follows.  First, the authors of ref.~\cite{Drummond:2006rz} noticed that integrals appearing in the perturbative computations of refs.~\cite{Anastasiou2003a, Bern2005} have a curious inversion property in the dual space.  Together with the obvious Lorentz symmetry, this generates the conformal group. This symmetry was then confirmed, and in fact used to guide the computations, at higher loop orders and for larger numbers of external particles in refs.~\cite{Bern2007, Bern2007a, Bern2008b}.  In a parallel development~\cite{Alday2007a}, Alday and Maldacena showed how to compute scattering amplitudes in the dual string theory.  This turned out to be closely related to the computation of a Wilson loop (in a language more familiar to mathematicians, a Wilson loop is the trace of the holonomy of the connection \(A\) around a curve).  The strong coupling computation leads us to believe that there is a connection between scattering amplitudes and a Wilson loop around a polygonal contour with vertices \(x_i\).  This was confirmed also at weak coupling in several papers~\cite{Drummond2008c, Brandhuber2008a, Drummond2008b, Bern2008a, Drummond:2008aq}.  Under the duality the scattering amplitudes map to Wilson loops and the dual conformal symmetry of scattering amplitudes maps to the conformal symmetry of the Wilson loops.  Ref.~\cite{Drummond:2008vq} showed that in fact the scattering amplitudes enjoy a dual \emph{super}-conformal symmetry. This corresponds in the dual side to the superconformal symmetry of a Wilson super-loop, which is the trace of the holonomy of a superconnection in superspace along a polygonal contour.  The corresponding super-loops were first defined in refs.~\cite{Mason2010, Caron-Huot2011}.

The dual space \(\tilde{\mathbb{M}}\) is noncompact and it does not have an action of the conformal group since some points are sent to infinity under conformal transformations.  This problem can be solved by compactifying \(\tilde{\mathbb{M}}\) is a way compatible with the action of the conformal group.  Moreover, \(\tilde{\mathbb{M}}\) comes with a Minkowski signature.  It is more convenient to use complex coordinates instead and to impose reality conditions when needed. Doing this, we can treat both the cases of Lorentz signature and of split signature.  The complexified and compactified dual space can be represented as the \(\mathbb{G}(2,4)\) Grassmannian of two-planes in \(\mathbb{C}^{4}\) containing the origin.  Therefore, to each point in dual space \(\tilde{\mathbb{M}}\) we can associate a two-plane in \(\mathbb{C}^{4}\).  Two points in dual space are light-like separated if their corresponding planes intersect in a line (it is easy to check that this imposes one constraint).  If we projectivize this construction, to a line through the origin in \(\mathbb{C}^{4}\) corresponds a point in \(\mathbb{CP}^{3}\) and to a two-plane through the origin in \(\mathbb{C}^4\) corresponds a projective line in \(\mathbb{CP}^3\).  We can do this for all pairs of points \((x_{i-1}, x_{i})\) and associate to each of them a point \(Z_{i} \in \mathbb{CP}^{3}\).  So instead of describing the kinematics by giving the momenta \(p_{i}\) subject to on-shell conditions \(p_{i}^{2} = 0\) and momentum conservation \(\sum_{i=1}^{n} p_{i} = 0\), we can describe it by giving \(n\) points \(Z_{i} \in \mathbb{CP}^{3}\).  The variables \(Z_{i}\) are known as momentum twistors\footnote{A similar construction can be done for Minkowski space \(\mathbb{M}\) instead, in which case we obtain the Penrose's twistor space (see ref.~\cite{Penrose:1967wn}).} and were introduced in ref.~\cite{Hodges:2009hk}.  Unlike for the variables \(p_{i}\) or \(x_{i}\), the momentum twistors are unconstrained.

The complexified dual conformal group acts as \(\grp{SL}(4, \mathbb{C})\) on the momentum twistors \([Z] \to [M Z]\), where \(M\) is an \(\grp{SL}(4, \mathbb{C})\) matrix and we have denoted by \([Z]\) the homogeneous coordinates of the point \(Z\).  The \(\grp{SL}(4, \mathbb{C})\) is the double cover of the complexified orthogonal group \(\grp{SO}(6, \mathbb{C})\).  There is a small subtlety here.  We defined the Lorentz group to be \(\grp{O}(1,3)\) and its complexification is \(\grp{O}(4, \mathbb{C})\).  However, the parity transformation in \(\grp{O}(4, \mathbb{C})\) does not embed in \(\grp{SO}(6, \mathbb{C})\), nor in its double cover \(\grp{SL}(4, \mathbb{C})\).  Then, the question is how does this discrete parity transformation act on the momentum twistor space.  The answer is as follows.  There is another space which, for lack of a better name, we call conjugate momentum twistor space whose points we label by \(W_i\).  There is a pairing between points in these two spaces, defined up to rescaling which we denote by \(W \cdot Z\).  Then we impose the rescaling invariant constraints \(W_{i} \cdot Z_{i} = 0\) and \(W_{i-1} \cdot Z_{i} = W_{i+1} \cdot Z_{i} = 0\) (here \(i \pm 1\) are considered modulo \(n\), the number of particles in the scattering process).  Given the \(Z_i\), the \(W_i\) are determined up to a rescaling.  Then, parity acts as the discrete transformation \(Z_{i} \leftrightarrow W_{i}\).

The translation of the kinematics to momentum twistor language makes it easy to build conformal invariants.  In order to make \(\grp{SL}(4, \mathbb{C})\) invariants, we can form four-brackets \(\langle i j k l\rangle = \text{Vol}(v_{i}, v_{j}, v_{k}, v_{l})\), where \(v_{i}\) is a vector in \(\mathbb{C}^{4}\) corresponding to \(Z_{i}\) and \(\text{Vol}\) is a volume form which is preserved by the action of \(\grp{SL}(4, \mathbb{C})\).

So we have established that we can describe the kinematics of a scattering process by giving a configuration of \(n\) ordered points \(Z_i\) in \(\mathbb{CP}^3\).  The homogeneous coordinates of these points fit in a \(4 \times n\) matrix.  The conformal invariants are built from the \(4 \times 4\) minors of this \(4 \times n\) matrix.

The description above is very similar to the description of coordinates on a Grassmannian.  For \(k \leq n\), the Grassmannian \(\mathbb{G}(k,n)\) of \(k\)-planes in an \(n\)-dimensional space can be described as the space of \(k \times n\) matrices of full rank modulo the left action by \(\grp{GL}(k)\).  Given such a \(k \times n\) matrix, we can form \(\binom{n}{k}\) minors of type \(k \times k\).  They can be labeled by \(k\) integers \(i_{1}, \dotsc, i_{k} \in \lbrace 1, \dotsc, n\rbrace\), corresponding to the columns of the initial \(k \times n\) matrix.  We will denote the determinants of these minors by \(\langle i_{1}, \dotsc, i_{k}\rangle\).  These determinants are also known as Pl\"ucker coordinates, and satisfy Pl\"ucker relations
\begin{equation}
  \label{eq:plucker-rel}
  \langle i, k, I\rangle \langle j, l, I\rangle = \langle i, j, I\rangle \langle k, l, I\rangle + \langle j, k, I\rangle \langle i, l, I\rangle,
\end{equation} where \(I\) is a multi-index with \(k-2\) entries.  The Pl\"ucker relations define an embedding, called Pl\"ucker embedding, of the Grassmannian into a projective space of dimension \(\binom{n}{k}\).

In the next section we will show that the Pl\"ucker relations in eq.~(\ref{eq:plucker-rel}) are the same as the exchange relations in a cluster algebra (see eq.~(\ref{eq:mutation}, for example).  This will also provide a way to build more complicated coordinates starting from simple minors.  Such combinations naturally appear in expressions for scattering amplitudes in \(\mathcal{N} = 4\).

Grassmannians have the important property of duality which identifies \(\mathbb{G}(k, n)\) with \(\mathbb{G}(n-k, n)\).  This is useful since it allows to simplify the geometric picture (as has been done in refs.~\cite{Goncharov2010, Golden:2013xva}).  Consider first the case \(n=6\).  The kinematics is described by a configuration of six ordered points in \(\mathbb{CP}^3\) or by the Grassmannian \(\mathbb{G}(4,6)\).  By Grassmannian duality this is the same as \(\mathbb{G}(2,6)\) which then can be translated to a configuration of six ordered points in \(\mathbb{CP}^1\), a much simpler-looking (though equivalent) geometric configuration.

A similar simplification can be performed for the case of \(n=7\), where a configuration of seven points in \(\mathbb{CP}^3\) can be mapped to a configuration of seven points in \(\mathbb{CP}^2\).  In general, this means that the configurations of \(n\) ordered points in \(\mathbb{CP}^{k-1}\) are the same as configurations of \(n\) ordered points in \(\mathbb{CP}^{n-k-1}\).  Therefore we can restrict to \(2 \leq k \leq \lfloor\tfrac {n-1} 2\rfloor\) without loss of generality.

\section{Introduction to cluster algebras}
\label{sec:intr-clust-algebr}

In this section we present some useful facts about cluster algebras. In the next section we will make the connection with Grassmannians and Pl\"ucker coordinates.  Cluster algebras have been introduced in a series of papers~\cite{1021.16017, 1054.17024, 1135.16013, 1127.16023} by Fomin and Zelevinsky.

Since the formal definition is a bit complicated, we will content ourselves with an informal description.  Cluster algebras are characterized as follows: they are commutative algebras constructed from distinguished generators (called \emph{cluster variables}) which are grouped into non-disjoint sets of constant cardinality (called \emph{clusters}).  The clusters are constructed recursively by an operation called \emph{mutation} from an initial cluster.  The number of variables in a cluster is called the rank of the cluster algebra.

Let us consider an example.  The \(A_{2}\) cluster algebra is defined by the following data:
\begin{itemize}
\item cluster variables: \(x_{m}, \quad m \in \mathbb{Z}\)
\item clusters: \(\lbrace x_{m}, x_{m+1}\rbrace\)
\item initial cluster: \(\lbrace x_{1}, x_{2}\rbrace\)
\item rank: \(2\)
\item exchange relations: \(x_{m-1} x_{m+1} = 1 + x_{m}\)
\item mutation: \(\lbrace x_{m-1}, x_{m}\rbrace \to \lbrace x_{m}, x_{m+1}\rbrace\).
\end{itemize}

Using the exchange relations we find that
\begin{equation}
 x_{3} = \frac {1+x_{2}}{x_{1}},\quad
 x_{4} = \frac {1+x_{1}+x_{2}}{x_{1} x_{2}},\quad
 x_{5} = \frac {1+x_{1}}{x_{2}},\quad
 x_{6} = x_{1}, \quad
 x_{7} = x_{2}, \quad \dots .
\end{equation}  Therefore, the sequence \(x_{m}\) is periodic with period five and the number of cluster variables is finite.

When expressing the cluster variables \(x_{m}\) in terms of the variables \((x_{1}, x_{2})\), we encounter two unexpected features (which hold in general for arbitrary cluster algebras).  First, the denominators of the cluster variables are always monomials.  In general, we expect the cluster variables to be rational fractions of the initial cluster variables, but in fact the denominator is always a monomial.  This is known under the name of ``Laurent phenomenon'' (see.~\cite{1021.16017}).  The second observation is that the numerator is a polynomial with positive coefficients.

As we alluded to before, this construction has a connection with Pl\"ucker
relations.  If we set \(x_1 = \tfrac {\langle 23\rangle \langle 14\rangle}{\langle 12\rangle \langle 34\rangle}\) and \(x_2 = \tfrac {\langle 13\rangle \langle 45\rangle}{\langle 34\rangle \langle 15\rangle}\), where \(\langle i j\rangle\) are coordinates of the Grassmannian \(\mathbb{G}(2,5)\), we can compute the rest of cluster variables by using the Pl\"ucker identities \(\langle i k\rangle \langle j l\rangle = \langle i j\rangle \langle k l\rangle + \langle i l\rangle \langle j k\rangle\), to obtain
\begin{gather*}
  x_1 = \frac {\langle 23\rangle \langle 14\rangle}{\langle 12\rangle \langle 34\rangle},\quad
  x_2 = \frac {\langle 13\rangle \langle 45\rangle}{\langle 34\rangle\langle 15\rangle},\quad
  x_3 = \frac {\langle 12\rangle\langle 35\rangle}{\langle 15\rangle\langle 23\rangle},\quad
  x_4 = \frac {\langle 25\rangle\langle 34\rangle}{\langle 23\rangle\langle 45\rangle},\quad
  x_5 = \frac {\langle 15\rangle\langle 24\rangle}{\langle 12\rangle\langle 45\rangle}.
\end{gather*}

In the following we will use a description of cluster algebras starting with quiver.  We now describe how to obtain a cluster algebra from a quiver.  A quiver is an oriented graph which we will require to be connected, finite, without loops (arrows with the same origin and target) and two-cycles (pairs of arrows going in opposite directions between two vertices).

Starting with a quiver with a given vertex \(k\) we define a new quiver obtained by mutating at vertex \(k\).  The new quiver is obtained by applying the following operations on the initial quiver:
\begin{itemize}
\item for each path \(i \to k \to j\) we add an arrow \(i \to j\),
\item reverse all the arrows on the edges incident with \(k\),
\item remove all the two-cycles that may have formed.
\end{itemize}  The mutation at \(k\) is an involution; when applied twice in succession we obtain the initial cluster.

Each quiver of the restricted type defined above is in one-to-one correspondence with skew-symmetric matrices, once we fix an ordering of the vertices.  The skew-symmetric matrix \(b\) is such that \(b_{i j}\) is the difference between the number of arrows \(i \to j\) and the number of arrows \(j \to i\).  Since only one of the terms above is nonvanishing, \(b_{i j} = -b_{j i}\).  Under a mutation at vertex \(k\) the matrix \(b\) transforms to \(b'\) given by
\begin{equation}
  \label{eq:b-mutation}
  b'_{i j} =
  \begin{cases}
    -b_{i j}, &\quad \text{if \(k \in \lbrace i, j\rbrace\)},\\
    b_{i j}, &\quad \text{if \(b_{i k} b_{k j} \leq 0\)},\\
    b_{i j} + b_{i k} b_{k j}, &\quad \text{if \(b_{i k}, b_{k j} > 0\)},\\
    b_{i j} - b_{i k} b_{k j}, &\quad \text{if \(b_{i k}, b_{k j} < 0\)}
  \end{cases}.
\end{equation}

If we start with a quiver with \(n\) vertices and associate to each vertex \(i\) a variable \(x_{i}\), we can use the skew-symmetric matrix \(b\) to define a mutation relation at the vertex \(k\) by
\begin{equation}
  \label{eq:mutation}
  x_{k} x_{k}' = \prod_{i \vert b_{i k} > 0} x_{i}^{b_{i k}} + \prod_{i \vert b_{i k} < 0} x_{i}^{-b_{i k}},
\end{equation} with the understanding that an empty product is set to one.  The mutation at \(k\) changes \(x_{k}\) to \(x_{k}'\) defined by eq.~(\ref{eq:mutation}) and leaves the other cluster variables unchanged.

The \(A_{2}\) cluster algebra can be expressed by a quiver \(x_{1} \to x_{2}\).  Then, a mutation at \(x_{1}\) replaces it by \(x_{1}' = \frac {1+x_{2}}{x_{1}} \equiv x_{3}\) and reverses the arrow.  A mutation at \(x_{2}\) replaces it by \(x_{2}' = \frac {1+x_{1}}{x_{2}} \equiv x_{5}\).  In the diagram~\eqref{eq:pentagon} below we represent the quivers and the mutations for the \(A_2\) cluster algebra (the arrows between quivers are labeled by the mutated variable).
\begin{equation}
\label{eq:pentagon}
\begin{xy} 0;<1pt,0pt>:<0pt,-1pt>::
  (100,30) *+{\framebox{$x_3 \leftarrow x_2$}} ="0",
  (85,70) *+{\framebox{$x_3 \to x_4$}} ="1",
  (10,70)*+{\framebox{$x_5 \leftarrow x_4$}} ="2",
  (0,30) *+{\framebox{$x_5 \to x_1$}} ="3",
  (50,0) *+{\framebox{$x_1 \to x_2$}} ="4",
  "0", {\ar^{x_2} "1"},
  "4", {\ar^{x_1} "0"},
  "1", {\ar^{x_3} "2"},
  "2", {\ar^{x_4} "3"},
  "3", {\ar^{x_5} "4"},
\end{xy}
\end{equation}

\section{The cluster algebra for \texorpdfstring{\(\mathbb{G}(k,n)\)}{G(k,n)}}
\label{sec:clust-algebra}

The Grassmannian \(\mathbb{G}(k,n)\) has a cluster algebra structure which was described in ref.~\cite{MR2078567} (this construction is also reviewed in ref.~\cite{1215.16012}).

For \(k < n\) we consider the description of the Grassmannian \(\mathbb{G}(k,n)\) as the equivalence classes of \(k \times n\) matrices of full rank, where two matrices are equivalent if they differ by the left action of a \(\grp{GL}(k)\) matrix.  If the leftmost \(k \times k\) minor is non-singular, i.e.\ \(\langle 1, \dotsc, k\rangle \neq 0\) then, by left multiplication with an appropriate \(\grp{GL}(k)\) matrix, we can transform it to the identity matrix.  After this operation the representative \(k \times n\) matrix has the form \((\mathbf{1}_{k}, Y)\), where \(\mathbf{1}_{k}\) is the \(k \times k\) identity matrix and \(Y\) is a \(k \times l\) matrix with \(l = n-k\).  The entries \(y_{i j}\), \(1 \leq i \leq k\), \(1 \leq j \leq l\) of the matrix \(Y\) are coordinates on the cell of the Grassmannian where \(\langle 1, \dotsc, k\rangle \neq 0\).

Now we define a matrix \(F_{i j}\) for \(1 \leq i \leq k\), \(1 \leq j \leq l\), which is the biggest square matrix which fits inside \(Y\) and whose lower-left corner is at position \((i,j)\) inside \(Y\). Then we define \(l(i,j) = \min(i-1, n-j-k)\) and
\begin{equation}
  f_{i j} = (-1)^{(k-i)(l(i,j)-1)} \det F_{i j}.
\end{equation}

According to ref.~\cite{MR2078567}, the initial quiver for the \(\mathbb{G}(k,n)\) cluster algebra is given by\footnote{Here we are presented a flipped version of the quiver and with the arrows reversed with respect to the quivers of refs.~\cite{MR2078567, 1215.16012}.}
\begin{equation}
  \label{eq:initial-quiver-gkn}
  \begin{xy} 0;<1pt,0pt>:<0pt,-1pt>::
(0,0) *+{f_{1 l}} ="0",
(50,0) *+{\cdots} ="1",
(100,0) *+{f_{13}} ="2",
(150,0) *+{f_{12}} ="3",
(200,0) *+{\framebox[5ex]{$f_{11}$}} ="4",
(0,50) *+{f_{2 l}} ="5",
(50,50) *+{\cdots} ="6",
(100,50) *+{f_{23}} ="7",
(150,50) *+{f_{22}} ="8",
(200,50) *+{\framebox[5ex]{$f_{21}$}} ="9",
(0,100) *+{\vdots} ="10",
(50,100) *+{\vdots} ="11",
(100,100) *+{\vdots} ="12",
(150,100) *+{\vdots} ="13",
(200,100) *+{\vdots} ="14",
(0,150) *+{\framebox[5ex]{$f_{kl}$}} ="15",
(50,150) *+{\cdots} ="16",
(100,150) *+{\framebox[5ex]{$f_{k3}$}} ="17",
(150,150) *+{\framebox[5ex]{$f_{k2}$}} ="18",
(200,150) *+{\framebox[5ex]{$f_{k1}$}} ="19",
"0", {\ar"1"},
"0", {\ar"5"},
"6", {\ar"0"},
"1", {\ar"2"},
"2", {\ar"3"},
"2", {\ar"7"},
"8", {\ar"2"},
"3", {\ar"4"},
"3", {\ar"8"},
"9", {\ar"3"},
"5", {\ar"6"},
"5", {\ar"10"},
"6", {\ar"7"},
"7", {\ar"8"},
"7", {\ar"12"},
"13", {\ar"7"},
"8", {\ar"9"},
"8", {\ar"13"},
"14", {\ar"8"},
"10", {\ar"15"},
"12", {\ar"17"},
"19", {\ar"13"},
"18", {\ar"12"},
"13", {\ar"18"},
\end{xy}
\end{equation}

The quiver above has two types of vertices, boxed and unboxed.  The boxed vertices are special and called \emph{frozen vertices}.  We do not allow mutations in the frozen vertices.  The associated variables to the frozen vertices are called \emph{coefficients} instead of \emph{cluster variables}.  We define the \emph{principal part} of such a quiver to be the quiver obtained by erasing the frozen vertices and the edges incident with them.

For the case \(n=5\) and \(k=2\), we can compute \(f_{11} = \langle 23\rangle\), \(f_{12} = \langle 24\rangle\), \(f_{13} = \langle 25\rangle\), \(f_{21} = \langle 34\rangle\), \(f_{22} = \langle 45\rangle\), \(f_{23} = \langle 15\rangle\).  Then, the the initial quiver diagram looks like below
\begin{equation}
\label{eq:g25}
\begin{xy} 0;<1pt,0pt>:<0pt,-1pt>::
(25,25) *+{25} ="0",
(75,25) *+{24} ="1",
(125,25) *+{\framebox[5ex]{$\langle 23\rangle$}} ="2",
(125,75) *+{\framebox[5ex]{$\langle 34\rangle$}} ="3",
(75,75) *+{\framebox[5ex]{$\langle 45\rangle$}} ="4",
(25,75) *+{\framebox[5ex]{$\langle 15\rangle$}} ="5",
(0,0) *+{\framebox[5ex]{$\langle 12\rangle$}} ="6",
"0", {\ar"1"},
"4", {\ar"0"},
"0", {\ar"5"},
"6", {\ar"0"},
"1", {\ar"2"},
"3", {\ar"1"},
"1", {\ar"4"},
\end{xy}
\end{equation}
where we have also included explicitly a frozen variable \(\langle 12\rangle\) which is equal to unity in the special parametrization we chose (on the part of the Grassmannian where \(\langle 12\rangle \neq 0\)).

After doing a mutation on the node \(\langle 14\rangle\), we obtain a similar quiver diagram where the frozen vertex \(\langle 15\rangle\) is special instead of \(\langle 34\rangle\).  Just like in the four-point case the arrows containing the mutated node get reversed and the link between \(\langle 13\rangle\) and \(\langle 34\rangle\) gets deleted and replaced with a link \(\langle 13\rangle \to \langle 15\rangle\).  It is easy to see that by mutating one gets the five similar quivers and nothing more.

The principal part of the quiver for configurations of five points in \(\mathbb{CP}^{1}\) is the same as the Dynkin diagram of \(A_{2}\) Lie algebra.  Indeed, this \emph{is} the \(A_2\) cluster algebra we discussed in sec.~\ref{sec:intr-clust-algebr}.  The appearance of the \(A_{2}\) Dynkin diagram provides the motivation for the name.  We can define scaling invariant cross-ratios associated to any unfrozen node by taking the ratio of the product of coordinates in the quiver which can be reached by going against the arrows going in by the product of coordinates in the quiver which can be reached by following the arrows going out.   For example, the cross-ratio corresponding to \(\langle 13\rangle\) in the quiver~\eqref{eq:g25} is given by \(\tfrac {\langle 12\rangle \langle 34\rangle}{\langle 14\rangle \langle 23\rangle}\). A mutation reverses the arrows and therefore transforms these ratios to their inverse.  These cross-ratios are the cluster variables of the \(A_2\) algebra, and the exchange relations following from the quiver description can be shown to be the same as the exchange relations of the \(A_2\) algebra.

More complicated cases appear for six points in \(\mathbb{CP}^{2}\), where we obtain a \(D_{4}\) Dynkin diagram.  We can start with an initial quiver at the left below and mutate at vertex \(\langle 236\rangle\) to obtain the principal part of the quiver shown at right, which is the same as the Dynkin diagram of \(D_{4}\).
\begin{equation}
  \label{eq:d4}
  \begin{xy} 0;<1pt,0pt>:<0pt,-1pt>::
(30,30) *+{\langle 236\rangle} ="0",
(30,80) *+{\langle 136\rangle} ="1",
(30,130) *+{\framebox[7ex]{$\langle 126\rangle$}} ="2",
(80,130) *+{\framebox[7ex]{$\langle 156\rangle$}} ="3",
(80,80) *+{\langle 356\rangle} ="4",
(80,30) *+{\langle 235\rangle} ="5",
(130,30) *+{\framebox[7ex]{$\langle 234\rangle$}} ="6",
(130,80) *+{\framebox[7ex]{$\langle 345\rangle$}} ="7",
(130,130) *+{\framebox[7ex]{$\langle 456\rangle$}} ="8",
(0,0) *+{\framebox[7ex]{$\langle 123\rangle$}} ="9",
"0", {\ar"1"},
"4", {\ar"0"},
"0", {\ar"5"},
"9", {\ar"0"},
"1", {\ar"2"},
"3", {\ar"1"},
"1", {\ar"4"},
"4", {\ar"3"},
"5", {\ar"4"},
"4", {\ar"7"},
"8", {\ar"4"},
"5", {\ar"6"},
"7", {\ar"5"},
\end{xy} \qquad \qquad
\begin{xy} 0;<1pt,0pt>:<0pt,-1pt>::
(30,30) *+{\bullet} ="0",
(30,80) *+{\bullet} ="1",
(80,80) *+{\bullet} ="4",
(80,30) *+{\bullet} ="5",
"1", {\ar"0"},
"0", {\ar"4"},
"5", {\ar"0"},
\end{xy}
\end{equation}

We should note that for the quiver in~\eqref{eq:d4}, the cross-ratio corresponding to the entry \(\langle 356\rangle\) is given by \(\tfrac {\langle 136\rangle \langle 235\rangle \langle 456\rangle}{\langle 156\rangle \langle 236\rangle \langle 345\rangle}\).  This is more complicated than the cross-ratios which were obtained previously and it has some interesting properties.  It appeared already in~\cite{Goncharov1995} (before the cluster algebras were discovered), in connection with functional equations for the trilogarithm.  For a geometrical interpretation of this quantity see sec.~\ref{sec:projective-geometry} and figs.~\ref{fig:triple1}, \ref{fig:triple2}, \ref{fig:triple3}.

In ref.~\cite{1054.17024}, Fomin and Zelevinsky showed that a cluster is of finite type (i.e.\ it has a finite number of cluster variables), if the principal part of its quiver can be transformed to a Dynkin diagram by a sequence of mutations.  Furthermore, if the principal part of the quiver contains a subgraph which is an affine Dynkin diagram, then the cluster algebra is of infinite type.  Using this characterization, one can show that the cluster algebras arising from \(\mathbb{G}(2,n)\) and \(\mathbb{G}(3,6)\), \(\mathbb{G}(3,7)\) and \(\mathbb{G}(3,8)\) are of finite type.  In ref.~\cite{1088.22009}, Scott has shown that all the other \(\mathbb{G}(k,n)\) with \(2 \leq k \leq \tfrac n 2\) are of infinite type.

This has striking implications for scattering amplitudes in \(\mathcal{N}=4\) super-Yang-Mills theory which, as we have reviewed, are based on Grassmannians \(\mathbb{G}(4,n)\), for \(n \geq 6\).  If \(n=6\) we obtain \(\mathbb{G}(4,6) = \mathbb{G}(2,6)\) which is of finite type.  If \(n=7\) we obtain \(\mathbb{G}(4,7) = \mathbb{G}(3,7)\) which is again of finite type.  However, starting at eight-point the cluster algebras are not of finite type anymore.

Notice that the seeds we have been using break the cyclic symmetry of the configuration of points.  In order to see that the cyclic symmetry is preserved we need to show that two quivers whose labels are permuted by one unit are linked by a sequence of mutations.  This can be shown in full generality (see ref.~\cite{Golden:2013xva} for details).

So far the most studied cases were \(\mathbb{G}(4,n)\) for \(n=6, 7\). The case \(n=8\) is more complicated also because the cluster algebra is infinite.  In the remainder of this section we will list a few of the cluster coordinates appearing for \(\mathbb{G}(4,8)\) and discuss their properties.  By using mutations, one encounters
\begin{equation}
  \langle 1 2 (3 4 5) \cap (6 7 8)\rangle \equiv \langle 1 3 4 5\rangle \langle 2 6 7 8\rangle - \langle 2 3 4 5\rangle \langle 1 6 7 8\rangle.
\end{equation}  Here, the \(\cap\) notation emphasizes the following geometrical fact: the composite bracket \(\langle 1 2 (3 4 5) \cap (6 7 8)\rangle\) vanishes whenever the projective line \((3 4 5) \cap (6 7 8)\) obtained by intersecting two projective planes \((3 4 5)\) and \((6 7 8)\) and the points \(1\) and \(2\) lie in the same projective plane.    This notation has been introduced in ref.~\cite{Arkani-Hamed2011}.

Already for \(n=7\) we encounter \(\langle 1 2 (3 4 5) \cap (5 6 7)\rangle\),  when expressed in \(\mathbb{CP}^{3}\) language.  In previous work (see ref.~\cite{Goncharov1995}) a different notation has been used for this quantity.  First, a transformation to \(\mathbb{CP}^2\) language was performed.  Points in \(\mathbb{CP}^2\) can be represented as vectors in \(\mathbb{C}^3\), modulo rescalings. For two three-vectors \(v_1\), \(v_2\) we have a notion of vector product \(v_1 \times v_2\) which is the vector orthogonal to the plane spanned by \(v_1\) and \(v_2\).  Then, the composite brackets containing \(\cap\) can be translated to
\begin{equation}
  \langle v_1 \times w_1, v_2 \times w_2, v_3 \times w_3\rangle =
  \langle v_1 v_2 w_2\rangle \langle w_1 v_3 w_3\rangle -
  \langle w_1 v_2 w_2\rangle \langle v_1 v_3 w_3\rangle.
\end{equation}
Above, the right-hand side does not have the same manifest symmetry as the left-hand side so more equivalent expressions can be found by applying permutations to the vector labels.  Notice that the left-hand side vanishes when \(v_1 \times w_1\) and \(v_2 \times w_2\) differ by a rescaling.  This is equivalent to the statement that the planes spanned by \((v_1, w_1)\) and \((v_2, w_2)\) are identical.  Hence, \(\langle v_1 v_2 w_2\rangle = 0\) and \(\langle v_2 w_1 w_2\rangle = 0\) so the right-hand side vanishes as well.

Since the \(\mathbb{G}(4,8)\) cluster algebra is infinite, we are bound to find more and more complicated expressions.  One remarkable feature of the mutations is that the denominator can always be canceled by the numerator, after using Pl\"ucker identities. Therefore, these coordinates always seem to be \emph{polynomials} in the Pl\"ucker coordinates.  This is an analog of the Laurent phenomenon, but this time we obtain polynomials.\footnote{This holds in many explicit examples, but I have not found a proof in the literature.}  As an example in \(\mathbb{G}(4,8)\), we have the following identity
\begin{equation}
  \frac {\langle 1 2 3 7\rangle \langle 1 2 4 5\rangle \langle 1 6 7 8\rangle + \langle 1 2 7 8\rangle \langle 4 5 (6 7 1) \cap (1 2 3)\rangle}{\langle 1 2 6 7\rangle} = \langle 4 5 (7 8 1) \cap (1 2 3)\rangle.
\end{equation}  Here the left-hand side is the expression obtained following a mutation, while the right-hand side is the expression where the denominator has been canceled.

Even more complicated coordinates can be generated.  As an example, we also find
\begin{equation}
  \langle (1 2 3) \cap (3 4 5), (5 6 7) \cap (7 8 1)\rangle.
\end{equation}  This vanishes when the lines \((1 2 3) \cap (3 4 5)\) and \((5 6 7) \cap (7 8 1)\) intersect.  Equivalently, we can say that the lines \((3 4 5) \cap (5 6 7)\) and \((7 8 1) \cap (1 2 3)\) intersect.

\section{Poisson brackets}
\label{sec:poisson-brackets}

One can define a Poisson bracket on the cluster coordinates.  It is enough to define the Poisson bracket between the coordinates in a given cluster.  If \(X_i\), \(X_j\) belong to the same cluster, i.e.\ they are vertices in the same quiver, then their Poisson bracket is defined as
\begin{equation}
  \label{eq:poisson-x-coords}
  \lbrace X_{i}, X_{j}\rbrace = b_{i j} X_{i} X_{j},
\end{equation} where \(b_{i j} = -b_{j i}\) is the \(b\) matrix of the cluster.  The Poisson bracket is compatible with mutations.  That is,
\begin{equation}
  \lbrace X_{i}', X_{j}'\rbrace = b_{i j}' X_{i}' X_{j}',
\end{equation} where \(X_{i}'\) and \(b_{i j}'\) are obtained by a mutation from \(X_{i}\) and \(b_{i j}\), respectively.

The Poisson structure is easiest to understand for \(\mathbb{G}(2,n)\) cluster algebras (see ref.~\cite{MR2567745} for a discussion).  To a configuration of \(n\) points in \(\mathbb{CP}^{1}\) with a cyclic ordering we associate a convex polygon.  Each of the vertices of this polygon corresponds to one of the \(n\) points.

Then consider a complete triangulation of the polygon.  Each of the \(n-3\) diagonals in this triangulation determines a quadrilateral and therefore four points in \(\mathbb{CP}^{1}\).  Suppose a diagonal \(E\) determines a quadrilateral with vertices \(i,j,k,l\) where the ordering is the same as the ordering of the initial polygon.  Using these four points we can form a cross-ratio \(r(i,j,k,l) = \frac {z_{i j} z_{k l}}{z_{j k} z_{i l}}\).  We have \(r(i,j,k,l) = r(k,l,i,j)\) which implies that the cross-ratio is uniquely determined by the diagonal \(E\) and we don't have to chose an orientation.

If we flip the diagonal \(E\) then the initial cross-ratio goes to its inverse, but the cross-ratios corresponding to neighboring quadrilaterals change in a more complicated way.  In fact, they transform in the same way as the cluster coordinates, if the matrix \(b_{i j}\) is defined as follows.  Two diagonals \(E\) and \(F\) in a given triangulation are called adjacent if they are the sides of one of the triangles of the triangulation.  If the diagonals are adjacent we set \(b_{E F} = 1\) if the diagonal \(E\) comes before \(F\) when listing the diagonals at the common vertex in clockwise order. Otherwise we set \(b_{E F} = -1\).  If two diagonals \(E\) and \(F\) are not adjacent we set \(\epsilon_{E F} = 0\).

In general, it is hard to compute the Poisson bracket between two coordinates in different clusters.  One approach is to express the second coordinate in terms of the coordinates of a cluster containing the first one.  Then, we can use the definition.  In general this is hard.  Another approach is to use the Sklyanin bracket (see ref.~\cite{MR2078567}).  To explain this, we restrict again to the part of the Grassmannian \(\mathbb{G}(k, n)\) where \(\langle 1, \dots, k\rangle \neq 0\) and we use a representative under the left \(\grp{GL}(k)\) action which is \((\mathbf{1}_k, Y)\), where \(Y\) is a \(k \times l\), \(l = n-k\) matrix.  We denote the entries of the matrix \(Y\) by \(y_{i j}\), \(i = 1, \dots k\), \(j = 1, \dots, l\). On these coordinates we introduce a bracket called Sklyanin bracket given by
\begin{equation}
  \label{eq:sklyanin}
  \{y_{i j}, y_{\alpha \beta}\}_S = (\sgn(\alpha - i) - \sgn(\beta - j)) y_{i \beta} y_{\alpha j}.
\end{equation}
In general, Sklyanin bracket is defined using an \(R\)-matrix, which is a solution of a modified classical Yang-Baxter equation (see ref.~\cite{MR2078567} for details).

Now, we can extend the Sklyanin bracket to arbitrary functions of the variables \(y\), in the usual way
\begin{equation}
  \{f, g\}_S = \sum_{i,j,\alpha,\beta} \frac {\partial f}{\partial y_{i j}} \{y_{i j}, y_{\alpha \beta}\}_S \frac {\partial g}{\partial y_{\alpha \beta}}.
\end{equation}
This bracket satisfies the Jacobi identity, as can be shown by direct computation, using the identity \(\sgn(x) \sgn(y) + \sgn(y) \sgn(z) + \sgn(z) \sgn(x) = -1\) for \(x+y+z=0\) and \(x y z \neq 0\).

The cluster coordinates can be expressed in terms of variables \(y\) and their bracket can be computed using the formula above.  As an example, consider the case of the \(A_2\) algebra again.  There we have the cluster coordinates
\begin{equation}
  X_1 = \frac {(12)(45)}{(15)(24)} = -\frac {y_{12} y_{23} - y_{13} y_{22}}{y_{12} y_{23}}, \qquad
  X_2 = \frac {(25)(34)}{(23)(45)} = \frac {y_{13} (y_{11} y_{22} - y_{12} y_{21})}{y_{11} (y_{12} y_{23} - y_{13} y_{22})}.
\end{equation}
The computation of the bracket \(\{X_1, X_2\}_S\) is a bit tedious, but straightforward.  We find
\begin{equation}
  \{X_1, X_2\}_S = 2 X_1 X_2.
\end{equation}
Up to a factor of \(2\), we obtain the answer expected from the definition in terms of the \(b\) matrix of the quiver.  Now, we can compute Poisson brackets between any cluster coordinates, even if they don't belong to the same cluster.  Most of the Poisson brackets between coordinates which don't belong to the same cluster will be very complicated, but sometimes one obtains zero.  This information combined with other physical requirements, can uniquely determine some parts of the amplitudes, as done for example in ref.~\cite{Golden2015}.

\section{Elements of projective geometry}
\label{sec:projective-geometry}

It is very useful to understand the cross-ratios geometrically.  For example, the \(A_2\) cluster algebra described above involves the geometry of five points on \(\mathbb{CP}^1\).

The simplest type of cross-ratio is the cross-ratio of four points \((a,b,c,d)\) in \(\mathbb{CP}^{1}\).  If the points have have coordinates \((z_{a}, z_{b}, z_{c}, z_{d})\), then their cross-ratio is
\begin{equation}
  r(a,b,c,d) = \frac {z_{a b} z_{c d}}{z_{b c} z_{d a}},
\end{equation}
with \(z_{a b} = z_a - z_b\).  In the following we will try to reduce more complicated situations to configurations of four points on a projective line.

\begin{figure}
  \centering
  \includegraphics{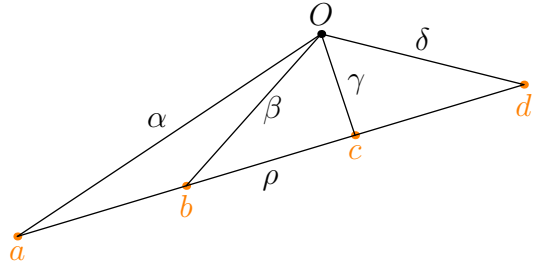}
  \caption{The cross-ratio of four lines in \(\mathbb{CP}^2\).}
  \label{fig:linesCR}
\end{figure}

By duality, a point in \(\mathbb{CP}^{2}\) is in correspondence with a line in \(\mathbb{CP}^{2}\).  A configuration of four points on a projective line in \(\mathbb{CP}^2\) dualizes to a configuration of four lines intersecting in a point.  Therefore, we can talk about the cross-ratio of four lines in \(\mathbb{CP}^{2}\) (see fig.~\ref{fig:linesCR}).

The cross-ratios of four lines \((\alpha, \beta, \gamma, \delta)\) containing a point \(O\) can be related to the cross-ratio of four points by taking an arbitrary line \(\rho\) (not containing the point \(O\)) and computing the intersection points \(a = \rho \cap \alpha\), \(b = \rho \cap \beta\), \(c = \rho \cap \gamma\), \(d = \rho \cap \delta\).  Then, the cross-ratio of the points \((a,b,c,d)\) on \(\rho\) is independent on \(\rho\) and is equal to the cross-ratio of the lines \((\alpha, \beta, \gamma, \delta)\)
\begin{equation}
  r(\alpha, \beta, \gamma, \delta) = r(a,b,c,d).
\end{equation}

\begin{figure}
  \centering
  \includegraphics{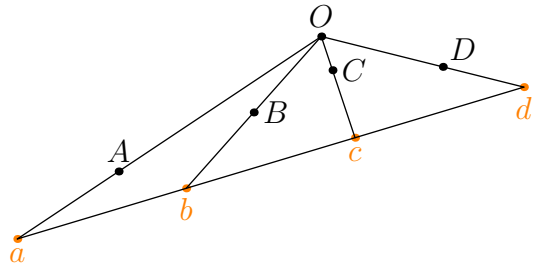}
  \caption{The cross-ratio of four lines determined by their common intersection point \(O\) and another point on each on of them.}
  \label{fig:projCR}
\end{figure}

If the lines are defined by pairs of points \(\alpha = (O A)\), \(\beta = (O B)\), \(\gamma = (O C)\), \(\delta = (O D)\), as in fig.~\ref{fig:projCR}, then the cross-ratio of the four lines is
\begin{equation}
  r(\alpha, \beta, \gamma, \delta) = r(a, b, c, d) = (O\vert A,B,C,D) \equiv \frac {\langle O A B\rangle \langle O C D\rangle}{\langle O B C\rangle \langle O D A\rangle},
\end{equation} where \(\langle X Y Z\rangle\) is proportional to the oriented area of the triangle \(\Delta(X,Y,Z)\).

\begin{figure}
  \centering
  \includegraphics{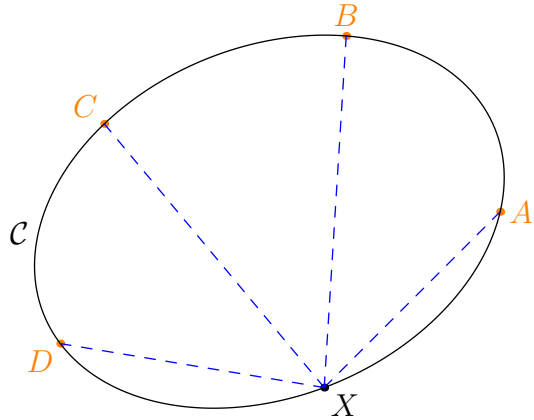}
  \caption{The cross-ratio of points \(A\), \(B\), \(C\), \(D\) with respect to the conic \(\mathcal{C}\).}
  \label{fig:ellipse}
\end{figure}

If the four points \(A\), \(B\), \(C\), \(D\) do not belong to a line we can't generically define their cross-ratio.  However, given a conic \(\mathcal{C}\) such that \(A\), \(B\), \(C\), \(D\) belong\footnote{Any conic is determined by five points.  Given four points there is an infinity of conics which contain them.} to \(\mathcal{C}\), then we can define their cross-ratio as follows: pick a point \(X\) on the conic \(\mathcal{C}\).  Then, by Chasles' theorem the cross-ratio of the lines \((X A)\), \((X B)\), \((X C)\) and \((X D)\) is independent on the point \(X\) and is defined to be the cross-ratio of the points \(A\), \(B\), \(C\), \(D\) (with respect to the conic \(\mathcal{C}\)).  See fig.~\ref{fig:ellipse}.

Let us now discuss the triple ratio of six points in \(\mathbb{CP}^{2}\) which was introduced by Goncharov.  We take the six points to be \(A\), \(B\), \(C\), \(X\), \(Y\), \(Z\). Numerically, this triple ratio is given by
\begin{equation}
  \label{eq:triple-ratio}
  r_{3}(A,B,C;X,Y,Z) = \frac {\langle ABX\rangle \langle BCY\rangle \langle CAZ\rangle}{\langle ABY\rangle \langle BCZ\rangle \langle CAX\rangle}.
\end{equation}

\begin{figure}
  \centering
  \includegraphics{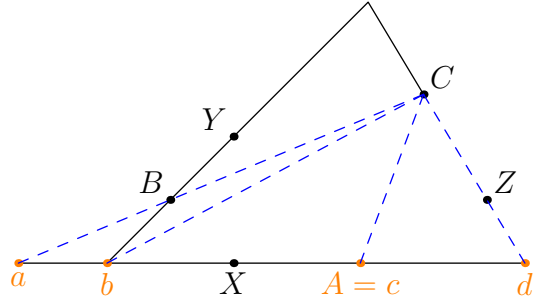}
  \caption{Triple ratio, expressed as a cross-ratio of points on the line \((AX)\).}
  \label{fig:triple1}
\end{figure}

It turns out that this ratio has several geometrical interpretations. Consider first the situation in fig.~\ref{fig:triple1}.  There, we have four lines which are dashed and blue: \(\alpha = (CB)\), \(\beta = (Cb)\), \(\gamma = (Cc)\), \(\delta = (Cd)\), where \(b = (AX) \cap (BY)\), \(c = A\) and \(d = (CZ) \cap (AX)\).  Their cross-ratio, obtained by intersecting with the line \((AX)\), is given by
\begin{equation}
  r(\alpha, \beta, \gamma, \delta) = r(a, b, c, d) = (C\vert B, (AX) \cap (BY), A, Z).
\end{equation}

\begin{figure}
  \centering
  \includegraphics{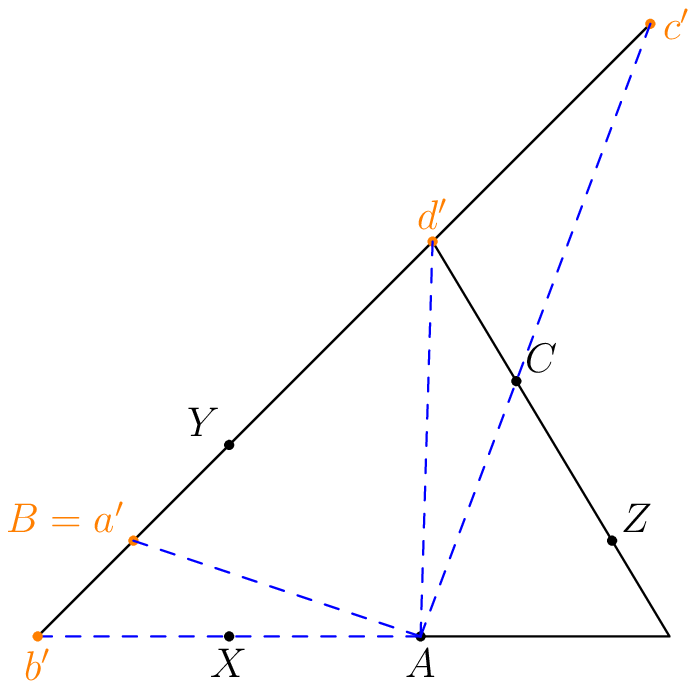}
  \caption{Triple ratio, expressed as a cross-ratio of points on the line \((BY)\).}
  \label{fig:triple2}
\end{figure}

But, instead of considering the intersections of the lines \((\alpha, \beta, \gamma, \delta)\) with the line \((AX)\) as above, we can consider the intersection with the line \((BY)\).  The intersection points are
\begin{align}
  a' &= \alpha \cap (BY) = B,\\
  b' &= \beta \cap (BY) = b = (AX) \cap (BY),\\
  c' &= \gamma \cap (BY) = (CA) \cap (BY),\\
  d' &= \delta \cap (BY) = (CZ) \cap (BY).
\end{align}  The corresponding figure is fig.~\ref{fig:triple2}.  If we denote by \(\alpha' = (AB)\), \(\beta' = (AX)\), \(\gamma' = (AC)\), \(\delta' = (Ad')\), we have
\begin{multline}
  r(a,b,c,d) = r(\alpha, \beta, \gamma, \delta) = r(a',b',c',d') =\\= r(\alpha', \beta', \gamma', \delta') = (A\vert B, X, C, (BY) \cap (CZ)).
\end{multline}

\begin{figure}
  \centering
  \includegraphics{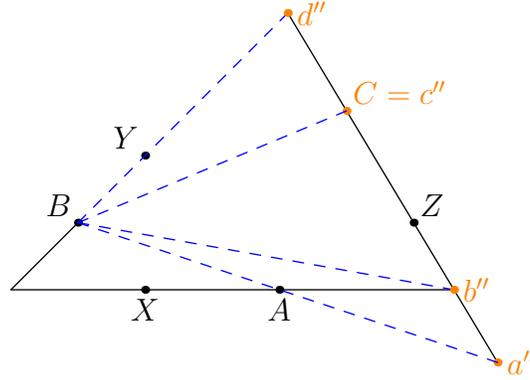}
  \caption{Triple ratio, expressed as a cross-ratio of points on the line \((CZ)\).}
  \label{fig:triple3}
\end{figure}

Now we can repeat the previous procedure.  We compute the cross-ratio \(r(\alpha', \beta', \gamma', \delta')\) by considering the intersection with \((CZ)\).  The intersection points are
\begin{align}
  a'' &= \alpha' \cap (C Z) = (A B) \cap (C Z),\\
  b'' &= \beta' \cap (C Z) = (A X) \cap (C Z),\\
  c'' &= \gamma' \cap (C Z) = C,\\
  d'' &= \delta' \cap (C Z) = (B Y) \cap (C Z).
\end{align}  See fig.~\ref{fig:triple3} for a geometrical representation.  If we define the lines \(\alpha'' = (B A)\), \(\beta'' = (B b'')\), \(\gamma'' = (B C)\), \(\delta'' = (B d'')\), we have
\begin{multline}
  (B\vert A, (C Z) \cap (A X), C, Y) = r(\alpha'', \beta'', \gamma'', \delta'') = r(a'', b'', c'', d'') = r(\alpha', \beta', \gamma', \delta').
\end{multline}

We have therefore shown that
\begin{equation}
  \label{eq:proj-equality}
  (A\vert B, X, C, (B Y) \cap (C Z)) = (B\vert A, (C Z) \cap (A X), C, Y) = (C\vert B, (A X) \cap (B Y), A, Z).
\end{equation}  Notice that this is also implied by the symmetry \(r_{3}(A,B,C;X,Y,Z) = r_{3}(B,C,A;Y,Z,X)\).

Let us now show that the invariant \((A\vert B, X, C, (B Y) \cap (C Z))\) has the same zeros and poles as \(r_{3}(A,B,C; X,Y,Z)\).  Form the definition, we know that \((A\vert B, X, C, (B Y) \cap (C Z))\) vanishes when \(\langle A B X\rangle = 0\) or \(\langle A C (B Y) \cap (C Z)\rangle = 0\).  The second three-bracket vanishes if \(\langle B C Y\rangle = 0\) or \(\langle C A Z\rangle = 0\).  In the first case \(B, C, Y\) are collinear and therefore \((B Y) \cap (C Z) = C\) so we have \(\langle A C (B Y) \cap (C Z)\rangle = \langle A C C\rangle = 0\).  In the second case, when \(\langle C A Z\rangle = 0\) we have that \(A \in (C Z)\), \(C \in (C Z)\) and \(P \equiv (B Y) \cap (C Z) \in (C Z)\).  Since all the entries of the three-bracket are collinear, we find that \(\langle A C (B Y) \cap (C Z)\rangle = 0\). We have shown that \((A\vert B, X, C, (B Y) \cap (C Z))\) vanishes if \(\langle A B X\rangle = 0\) or \(\langle B C Y\rangle = 0\) or \(\langle C A Z\rangle = 0\) which is the same as the numerator of \(r_{3}(A,B,C;X,Y,Z)\).  In order to find the poles we reason in the same way.

\section{Polylogarithm identities}
\label{sec:polylog-ident}

In this section we provide some more mathematical details on transcendental functions and explain how to partially integrate them. We denote by \(\mathcal{L}_n\) the Abelian group (under addition) of transcendental functions of transcendentality weight \(n\).  An important character in this story is the Bloch group \(B_n\), also called the classical polylogarithm group: it is the subgroup of \(\mathcal{L}_n\) generated by the classical polylogarithm functions \(\Li_n\) and their products.

Consider first the simplest kind of transcendental function, the logarithm.  If we are working modulo \(2 \pi i\), then we have that \(\ln z + \ln w = \ln (z w)\), for any \(z, w \in \mathbb{C}^*\).  In order to express this simple functional relation formally, define \(\mathbb{Z}[\mathbb{C}^{*}]\) to be the free Abelian group generated by \(\lbrace z\rbrace\), with integer coefficients and \(z\) non-zero complex numbers.  Concretely, elements of this group are quantities like \(\lbrace z\rbrace + \lbrace w\rbrace\) and the group operation is defined in the obvious way.  Then, we can quotient this group by the relations satisfied by the logarithm to obtain the logarithm group \(B_{1}\),
\begin{equation}
  B_{1} = \mathbb{Z}[\mathbb{C}^{*}]/(\lbrace z\rbrace + \lbrace w\rbrace - \lbrace z w\rbrace).
\end{equation}  This group is isomorphic to the multiplicative group of complex numbers, \(\mathbb{C}^{\times}\).

The next simplest transcendental functions are the dilogarithms, \(\Li_{2}\).  The dilogarithms satisfy a simple five-term functional relation.  One way to express this functional relation is to consider five points on \(\mathbb{CP}^{1}\) with coordinates \(z_{1}, \dotsc, z_{5}\).  From any four such points we can form a cross-ratio \(r(z_{1}, \dotsc, \hat{z}_{i}, \dotsc z_{5})\), where the hatted argument is missing. We use the definition \(r(i,j,k,l) = \tfrac {z_{i j} z_{kl}}{z_{j k} z_{l i}}\) with \(z_{ij} = z_i - z_j\).  Then the five-term identity can be written as
\begin{equation}
  \sum_{i=1}^{5} (-1)^{i} \Li_{2}(-r(z_{1}, \dotsc, \hat{z}_{i}, \dotsc, z_{5})) = \text{logs},
\end{equation} where we have denoted by \(\text{logs}\) the terms which can be written uniquely in terms of logarithms.  There is a theorem (see ref.~\cite{MR1760901}) that all the relations between dilogarithms are consequences of the five-term relations.  We can now define the Bloch group \(B_{2}\) by analogy to the logarithm case.  We first define \(\mathbb{Z}[\mathbb{C}]\) to be the free Abelian group generated by \(\lbrace z\rbrace_{2}\), where \(z\) is a complex number.  Then, we quotient be the five-term relations and the quotient is denoted by \(B_{2}\)
\begin{equation}
  B_{2} = \mathbb{Z}[\mathbb{C}]/(\text{five-term relations}).
\end{equation}  In this case we have a group morphism \(\delta\), \(B_{2} \xrightarrow{\delta} \Lambda^{2} \mathbb{C}^{*}\) which is defined by \(\delta(\lbrace z\rbrace_{2}) = (1-z) \wedge z\).  To check that this is a group morphism we need to show that \(\delta(\text{five-term relation}) = 0\) or
\begin{equation}
  \sum_{i=1}^{5} (-1)^{i} (1 + r(z_{1}, \dotsc, \hat{z}_{i}, \dotsc, z_{5})) \wedge r(z_{1}, \dotsc, \hat{z}_{i}, \dotsc, z_{5}) = 0,
\end{equation} which can be done by a short computation.

Let us now discuss \(\Li_{3}\) functions.  There is a theorem stating that all transcendentality three functions can be written as a linear combination of \(\Li_{3}\) and products of lower transcendentality functions (see ref.~\cite{Goncharov1995}).

Just like in the previous cases, we first need to find the functional relations satisfied by \(\Li_{3}\) functions.  The identity satisfied by \(\Li_{3}\) is very similar to the one satisfied by \(\Li_{2}\) and can be described in terms of configurations of seven points on \(\mathbb{CP}^{2}\).  It is convenient to describe each of these points in terms of their homogeneous \(v_{i} \in \mathbb{C}^{3}\) coordinates, with \(i=1, \dots, 7\).  For three such vectors \(v_{i}\), \(v_{j}\), \(v_{k}\) we can define a three-bracket \(\langle \cdot, \cdot, \cdot\rangle : \mathbb{C}^{3} \times \mathbb{C}^{3} \times \mathbb{C}^{3} \to \mathbb{C}\) by the volume of the parallelepiped generated by them \(\langle i,j,k\rangle = \text{Vol}(v_{i}, v_{j}, v_{k})\).

Given six points in \(\mathbb{CP}^{3}\), we can form a cross-ratio
\begin{equation}
  r_{3}(1,2,3,4,5,6) = \frac {\langle 124\rangle \langle 235\rangle \langle 316\rangle}{\langle 125\rangle \langle 236\rangle \langle 314\rangle}.
\end{equation}  Such cross-ratios have been introduced and extensively used in ref.~\cite{Goncharov1995} and we also discuss their geometric interpretation in sec.~\ref{sec:clust-algebra}.  The \(\Li_{3}\) functional relations can be expressed in terms of this cross-ratio as
\begin{equation}
  \sum_{i=1}^{7} (-1)^{i} \Alt_{6} \Li_{3}(- r_{3}(1, \dotsc, \hat{i}, \dotsc, 7)) \approx 0,
\end{equation} where \(\Alt_{6}\) mean antisymmetrization in the six points on which \(r_{3}\) depends and \(\approx\) means that we have omitted the terms which are products of lower transcendentality functions.

Now we define
\begin{equation}
  B_{3} = \mathbb{Z}[\mathbb{C}]/(\text{seven-term relations}).
\end{equation}
There is a morphism \(\delta : B_{3} \to B_{2} \otimes \mathbb{C}^{*}\), \(\delta(\lbrace x\rbrace_{3}) = \lbrace x\rbrace_{2} \otimes x\).  In order to show that this morphism is well-defined, we need to show that that \(\delta\) annihilates the seven-term relations.

It may seem that we can continue in the same way to higher transcendentality.  However, this is not the case.  At transcendentality four there are new functions which can not be expressed in terms of \(\Li_{4}\) and products of lower transcendentality functions.  We can define \(B_{n}\) for \(n \geq 4\) in the same way as before, but there is a bigger group \(\mathcal{L}_{n}\) which is the Abelian group related to weight \(n\) polylogs, some of which are not classical polylogs.

We defined \(B_n\) to be the Abelian groups generated by classical polylogs and \(\mathcal{L}_n\) to be the Abelian groups of all polylogs of weight \(n\).  Now we want to characterize them.  The most mathematically concise way to describe their (conjectural!) connection is by an exact sequence, which for \(n=4\) reads
\begin{equation}
  0 \to B_4 \to \mathcal{L}_4 \to \Lambda^{2} B_2 \to 0.
\end{equation}
An exact sequence is a sequence of maps between spaces such that the image of a map falls in the kernel of the next one.  In the example above, the first arrow says that \(B_4\) maps to \(\mathcal{L}_4\) injectively, which is obvious since \(B_4\) is contained in \(\mathcal{L}_4\).  The last arrow says that the map \(\mathcal{L}_4 \to \Lambda^{2} B_2\) is surjective.  This is less obvious, but it means that for any element of \(\Lambda^{2} B_2\) one can find a weight four polylog with that \(\Lambda^{2} B_2\) projection.

Finally, the rest of the sequence means that \(\ker(\mathcal{L}_4 \to \Lambda^{2} B_2) = B_4\).  This means that if a weight four polylog has zero \(\Lambda^{2} B_2\) projection, which is to say it belongs to \(\ker(\mathcal{L}_4 \to \Lambda^{2} B_2)\), then it is a classical polylog, and vice versa.

Notice that in fig.~\ref{fig:triple1}, we have five points \((a, b, X, c, d)\) on the line \((A X)\).  From five points \((z_{1}, \dotsc, z_{5})\) in \(\mathbb{CP}^{1}\) we can produce a dilogarithm identity
\begin{equation}
  \label{eq:dilog-identity}
  \sum_{i=1}^{5} (-1)^{i} \lbrace -r(z_{1}, \dotsc, \widehat{z_{i}}, \dotsc, z_{5})\rbrace_{2} = 0.
\end{equation}
This motivates us to find the expressions in terms of three-brackets for the other cross-ratios that can be constructed from these five points on \((A X)\) (see fig.~\ref{fig:triple1}):
\begin{align}
  r(b,X,A,d) &= \frac {\langle B X Y\rangle \langle A C Z\rangle}{\langle A \times X, B \times Y, C \times Z\rangle},\\
  r(a,X,A,d) &= (C\vert B,X,A,Z),\\
  r(a,b,A,d) &= r_{3}(A,B,C; X,Y,Z),\\
  r(a,b,X,d) &= r_{3}(X,B,C; A,Y,Z),\\
  r(a,b,X,A) &= (B\vert C,Y,X,A).
\end{align}

This provides a geometric proof for the following dilogarithm identity
\begin{multline}
  -\left\lbrace \frac {\langle B X Y\rangle \langle A C Z\rangle}{\langle A \times X, B \times Y, C \times Z\rangle}\right\rbrace_{2}
  +\left\lbrace \frac {\langle CBX\rangle \langle CAZ\rangle}{\langle CXA\rangle \langle CZB\rangle}\right\rbrace_{2}
  -\left\lbrace \frac {\langle ABX\rangle \langle BCY \rangle \langle CAZ\rangle}{\langle ABY\rangle \langle BCZ \rangle \langle CAX\rangle}\right\rbrace_{2}\\
  +\left\lbrace \frac {\langle XBA\rangle \langle BCY \rangle \langle CXZ\rangle}{\langle XBY\rangle \langle BCZ \rangle \langle CXA\rangle}\right\rbrace_{2}
  -\left\lbrace \frac {\langle BCY\rangle \langle BXA\rangle}{\langle BYX\rangle \langle BAC\rangle}\right\rbrace_{2} = 0.
\end{multline}

Here is a \(40\)-term trilogarithm identity which was discovered when analyzing results of two-loop computations in \(\mathcal{N} = 4\) theory
\begin{multline}
  \label{eq:40-term-li3}
  \left\lbrace -\frac {\langle 125\rangle\langle 134\rangle}{\langle 123\rangle\langle 145\rangle}\right\rbrace_3 +
  \left\lbrace -\frac {\langle 126\rangle\langle 145\rangle}{\langle 124\rangle\langle 156\rangle}\right\rbrace_3 +
  \left\lbrace -\frac {\langle 126\rangle\langle 145\rangle\langle 234\rangle}{\langle 123\rangle\langle 146\rangle \langle 245\rangle}\right\rbrace_3 +\\
  \frac 1 3 \left\lbrace -\frac {\langle 136\rangle\langle 145\rangle\langle 235}{\langle 123\rangle\langle 156\rangle\langle 345\rangle}\right\rbrace_3 +
  (\text{cyclic permutations}) -\\
  (\text{anti-cyclic permutations}) = 0.
\end{multline}

In order to check that the \(B_{2} \wedge \mathbb{C}^{*}\) projection of the \(40\)-term trilogarithm identity is zero we need some dilogarithm identities.  For example, one of the dilogarithm identities which is useful is
\begin{multline}
-\left\{-\frac{\langle 123\rangle  \langle 456\rangle }{\langle 1\times 2,3\times 4,5\times 6\rangle }\right\}_2-
\left\{-\frac{\langle 125\rangle \langle 134\rangle }{\langle 123\rangle \langle 145\rangle}\right\}_2-
\left\{-\frac{\langle 123\rangle \langle 156\rangle \langle 345\rangle }{\langle 125\rangle \langle 134\rangle \langle 356\rangle}\right\}_2+\\
\left\{-\frac{\langle 124\rangle \langle 156\rangle \langle 345\rangle }{\langle 125\rangle \langle 134\rangle \langle 456\rangle}\right\}_2-
\left\{-\frac{\langle 156\rangle \langle 345\rangle }{\langle 135\rangle  \langle 456\rangle }\right\}_2=0.
\end{multline}
It can be interpreted geometrically as five points \((3, 4, (15)\cap(34), (12)\cap(34), (34)\cap(56))\) on the line \((34)\).

The second useful dilogarithm identity is
\begin{multline}
  \left\{-\frac{\langle 156\rangle \langle 234\rangle}{\langle 1\times 2,3\times 4,5\times 6\rangle }\right\}_2
  -\left\{-\frac{\langle 136\rangle \langle 234\rangle}{\langle 123\rangle \langle 346\rangle} \right\}_2
  -\left\{-\frac{\langle 156\rangle \langle 236\rangle }{\langle 126\rangle \langle 356\rangle }\right\}_2\\
  +\left\{-\frac{\langle 123\rangle \langle 156\rangle \langle 346\rangle}{\langle 126\rangle \langle 134\rangle \langle 356\rangle }\right\}_2
  -\left\{-\frac{\langle 123\rangle \langle 256\rangle \langle 346\rangle }{\langle 126\rangle \langle 234\rangle \langle 356\rangle }\right\}_2 = 0.
\end{multline}
It can be interpreted geometrically as five points \((1, 2, (12)\cap(34), (12)\cap(36), (12)\cap(56))\) on the line \((12)\).

The third useful dilogarithm identity is
\begin{multline}
-\left\{-\frac{\langle 156\rangle  \langle 234\rangle }{\langle 1\times 2,3\times 4,5\times 6\rangle }\right\}_2+
\left\{-\frac{\langle 145\rangle \langle 234\rangle }{\langle 124\rangle  \langle 345\rangle}\right\}_2+
\left\{-\frac{\langle 156\rangle  \langle 245\rangle }{\langle 125\rangle \langle 456\rangle}\right\}_2-\\
\left\{-\frac{\langle 124\rangle \langle 156\rangle \langle 345\rangle }{\langle 125\rangle \langle 134\rangle \langle 456\rangle}\right\}_2+
\left\{-\frac{\langle 124\rangle \langle 256\rangle \langle 345\rangle }{\langle 125\rangle \langle 234\rangle \langle 456\rangle }\right\}_2 = 0.
\end{multline}
It can be interpreted geometrically as five points \((1, 2, (12)\cap(34), (12)\cap(45), (12)\cap(56))\) on the line \((12)\).

The fourth useful dilogarithm identity is
\begin{multline}
\left\{-\frac{\langle 123\rangle  \langle 456\rangle }{\langle 1\times 2,3\times 4,5\times 6\rangle }\right\}_2+
\left\{-\frac{\langle 125\rangle \langle 234\rangle }{\langle 123\rangle  \langle 245\rangle}\right\}_2+
\left\{-\frac{\langle 123\rangle \langle 256\rangle \langle 345\rangle }{\langle 125\rangle \langle 234\rangle \langle 356\rangle }\right\}_2-\\
\left\{-\frac{\langle 124\rangle \langle 256\rangle \langle 345\rangle }{\langle 125\rangle \langle 234\rangle \langle 456\rangle}\right\}_2+
\left\{-\frac{\langle 256\rangle  \langle 345\rangle }{\langle 235\rangle  \langle 456\rangle }\right\}_2=0.
\end{multline}
It can be interpreted geometrically as five points \((3, 4, (12)\cap(34), (25)\cap(34), (34)\cap(56))\) on the line \((34)\).

The identities above are the identities needed to show the vanishing of terms of type \(\ast \otimes \langle 123\rangle\) in the projection to \(B_{2} \otimes \mathbb{C}^{*}\) of the \(40\)-term trilogarithm identity.  For the terms of type \(\ast \otimes \langle 124\rangle\) the same identities are sufficient, but there is another, simpler identity too, written below
\begin{multline}
-\left\{-\frac{\langle 126\rangle \langle 145\rangle}{\langle 124\rangle \langle 156\rangle}\right\}_2+
\left\{-\frac{\langle 126\rangle \langle 245\rangle}{\langle 124\rangle \langle 256\rangle}\right\}_2-
\left\{-\frac{\langle 146\rangle \langle 245\rangle}{\langle 124\rangle \langle 456\rangle}\right\}_2+\\
\left\{-\frac{\langle 156\rangle \langle 245\rangle}{\langle 125\rangle \langle 456\rangle}\right\}_2-
\left\{-\frac{\langle 156\rangle \langle 246\rangle}{\langle 126\rangle \langle 456\rangle }\right\}_2=0.
\end{multline}

This identity is special because it does not depend on point \(3\) at all.  It can be more geometrically written as
\begin{equation}
  \left\lbrace (1\vert 2 6 5 4)\right\rbrace_{2}+
  \left\lbrace (2\vert 1 4 5 6)\right\rbrace_{2}+
  \left\lbrace (4\vert 1 6 5 2)\right\rbrace_{2}+
  \left\lbrace (5\vert 1 2 4 6)\right\rbrace_{2}+
  \left\lbrace (6\vert 1 5 4 2)\right\rbrace_{2} = 0.
\end{equation}

Curiously, this simple-looking identity has a slightly more obscure geometrical interpretation.  Through the five points \(1\), \(2\), \(4\), \(5\), \(6\) passes a unique conic \(\mathcal{C}\).  The cross-ratio \((1\vert 2 6 5 4)\) is the cross-ratio of the points \((2,6,5,4)\) with respect to the conic \(\mathcal{C}\).  But we can pick another point \(X \in \mathcal{C}\) and we have, by Chasles' theorem, that \((X\vert 2 6 5 4) = (1\vert 2 6 5 4)\).  Then the previous identity becomes
\begin{equation}
  \left\lbrace (X\vert 2 4 5 6)\right\rbrace_{2}-
  \left\lbrace (X\vert 1 4 5 6)\right\rbrace_{2}+
  \left\lbrace (X\vert 1 2 5 6)\right\rbrace_{2}-
  \left\lbrace (X\vert 1 2 4 6)\right\rbrace_{2}+
  \left\lbrace (X\vert 1 2 4 5)\right\rbrace_{2} = 0,
\end{equation}
which is the usual form of the dilogarithm identity, where the cross-ratios are cross-ratios of the lines \((X1)\), \((X2)\), \((X4)\), \((X5)\), \((X6)\).

\section{Open questions}
\label{sec:questions}

The scattering amplitudes in \(\mathcal{N} = 4\) theory split into sub-sectors which are not related by supersymmetry transformations. Scattering amplitudes in the simplest sectors are called MHV (maximally helicity violating) amplitudes, for historical reasons. More complicated sectors are called NMHV (next to MHV), etc.  The six-point MHV amplitude has transcendentality four but, surprisingly, can be expressed in terms of classical polylogarithms only, as found in ref.~\cite{Goncharov2010}.  The next simplest amplitudes are the six-point NMHV, or the seven point MHV, which can not be written in terms of classical polylogarithms, since their \(B_2 \wedge B_2\) projection does not vanish.

Consider the \(\Lambda^{2} B_{2}\) projection of the seven-point MHV amplitude computed in ref.~\cite{Caron-Huot2011a}.  In \(\mathbb{CP}^{2}\) language it is given by
\begin{multline}
  -\Big\lbrace -\frac {\langle 2 \times 3, 4 \times 6, 7 \times 1 \rangle}{\langle 1 6 7\rangle \langle 2 3 4\rangle}\Big\rbrace_{2} \wedge \Big\lbrace -\frac {\langle 7 \times 1, 2 \times 3, 4 \times 5\rangle}{\langle 1 2 7\rangle \langle 3 4 5\rangle}\Big\rbrace_{2}\\ %
  -\Big\lbrace -\frac {\langle 2 \times 3, 4 \times 6, 7 \times 1 \rangle}{\langle 1 6 7\rangle \langle 2 3 4\rangle}\Big\rbrace_{2} \wedge \Big\lbrace -\frac {\langle 2 3 4\rangle \langle 4 5 6\rangle}{\langle 2 4 6\rangle \langle 3 4 5\rangle}\Big\rbrace_{2}\\ %
  -\Big\lbrace -\frac {\langle 2 \times 3, 4 \times 6, 7 \times 1 \rangle}{\langle 1 6 7\rangle \langle 2 3 4\rangle}\Big\rbrace_{2} \wedge \Big\lbrace -\frac {\langle 1 4 6\rangle \langle 5 6 7\rangle}{\langle 1 6 7\rangle \langle 4 5 6\rangle}\Big\rbrace_{2}\\ %
  -\Big\lbrace -\frac {\langle 2 \times 3, 4 \times 6, 7 \times 1 \rangle}{\langle 1 6 7\rangle \langle 2 3 4\rangle}\Big\rbrace_{2} \wedge \Big\lbrace -\frac {\langle 5 \times 6, 7 \times 1, 2 \times 3\rangle}{\langle 1 2 3\rangle \langle 5 6 7\rangle}\Big\rbrace_{2}\\ %
  +\Big\lbrace -\frac {\langle 1 3 7\rangle \langle 4 6 7\rangle}{\langle 1 6 7\rangle \langle 3 4 7\rangle}\Big\rbrace_{2} \wedge \Big\lbrace -\frac {\langle 1 2 3\rangle \langle 3 4 7\rangle}{\langle 1 3 7\rangle \langle 2 3 4\rangle}\Big\rbrace_{2} %
  -\Big\lbrace -\frac {\langle 1 3 7\rangle \langle 4 6 7\rangle}{\langle 1 6 7\rangle \langle 3 4 7\rangle}\Big\rbrace_{2} \wedge \Big\lbrace -\frac {\langle 3 4 7\rangle \langle 4 5 6\rangle}{\langle 3 4 5\rangle \langle 4 6 7\rangle}\Big\rbrace_{2}\\ %
  + \text{cyclic permutations of \(1, 2, \dotsc, 7\)}.
\end{multline}
Goncharov suggested to look at the Poisson bracket \({x, y}\) for any \(\lbrace -x\rbrace_{2} \wedge \lbrace -y\rbrace_{2} \in \Lambda^{2} B_{2}\).  This is well-defined since \(\lbrace -x\rbrace_{2} \wedge \lbrace -y\rbrace_{2} = -\lbrace -y\rbrace_{2} \wedge \lbrace -x\rbrace_{2}\) and a similar sign change appears from the Poisson bracket.

It is not understood why, but we find that these Poisson brackets are zero.  We can show that for every term \(\lbrace -x\rbrace_{2} \wedge \lbrace -y\rbrace_{2} \in \Lambda^{2} B_{2}\) listed above there is at least one cluster containing \(x\) and \(y\).  In order to prove this, for every pair \((x,y)\) we need to exhibit a quiver graph which contains them and which is such that there are no arrows between \(x\) and \(y\).  Alternatively, one can compute the Sklyanin bracket as in sec.~\ref{sec:poisson-brackets}

As mentioned in the introduction, scattering amplitudes have the property of factorization (see ref.~\cite{Anastasiou2009}). Formulating this precisely and studying its implications for the cluster algebra structure would be very interesting.  A complete discussion would take us too far, but we want to mention only one important aspect: factorization only works if the transcendental functions satisfy some identities.

In mathematics one prefers to work with some \emph{real} analytic functions, like
\begin{gather}
  L_2(z) = \Im\left(\Li_2(z) + \ln |z| \ln(1-z)\right),\\
  L_3(z) = \Re\left(\Li_3(z) - \ln |z| \Li_2(z) - \frac 1 3 \ln^2 |z| \ln (1-z)\right),
\end{gather}
which have simple functional relations (modulo some additive constants, one can simply replace \(\{z\}_2 \to L_2(z)\) and \(\{z\}_3 \to L_3(z)\)) to obtain an identity for functions.  However, for physics we need to have \emph{complex} analytic functions instead. Therefore, it is not yet clear what are the best building blocks for the scattering amplitudes.

The reader might be puzzled by the following fact: we have a big symmetry group \(\grp{PSU}(2,2\vert 4)\) but in terms of Grassmannians only the conformal group \(\grp{SU}(2,2)\) or the complexified \(\grp{SL}(4)\) is visible.  How to make the rest of the symmetry visible?  This is not known at present.  Maybe recent developments like the definition of cluster superalgebras in ref.~\cite{Ovsienko2015} hold the key to further progress.

Are there other polylogarithm identities of cluster type?  As we have reviewed, the dilogarithm identity contains arguments which form an \(A_2\) (or \(\mathbb{G}(2,5)\) cluster algebra, while the trilogarithm identity contains arguments which form a \(D_4\) (or \(\mathbb{G}(3,6)\) cluster algebra.  A computer search for a \(\Li_4\) identity with arguments in finite cluster algebra did not find anything.  It is possible that there are such identities for infinite cluster algebras.

Before ending this brief review, let us point out some references which discuss complementary details.  Cluster algebras appeared in ref.~\cite{ArkaniHamed:2012nw} in connection with scattering amplitudes, but in a different way than we reviewed here. Ref.~\cite{Golden2014} also reviews the connection between scattering amplitudes and cluster algebras, with an emphasis on the combinatorics of Stasheff polytopes.  Ref.~\cite{Huang2014} reviews the case of a three-dimensional analog of the \(\mathcal{N}=4\) theory which we described here.

Many results were obtained by applying the bootstrap method (see refs.~\cite{Dixon:2011pw, Dixon2012a, Caron-Huot2012a, Dixon2013, Dixon2014, Golden2014a, Dixon2014a, Golden2015, Drummond2015}).

\section{Acknowledgments}

First, I would like to thank the organizers of the Opening Workshop of the Research Trimester on Multiple Zeta Values, Multiple Polylogarithms, and Quantum Field Theory: Jos\'e I. Burgos Gil, Kurusch Ebrahimi-Fard, D. Ellwood, Ulf K\"uhn, Dominique Manchon and P. Tempesta.

I would also like to thank the participants and particularly Fr\'ed\'eric Chapoton and Herbert Gangl for discussions during the opening workshop Numbers and Physics (NAP2014).  Finally, I am grateful to my coauthors in refs.~\cite{Goncharov2010, Golden:2013xva} for collaboration.

\bibliographystyle{plain}
\bibliography{talk}

\end{document}